\documentclass[hyper]{JHEP3_arxiv} 
\usepackage[utf8]{inputenc}
\usepackage{graphicx}
\usepackage{epsfig}
\usepackage{amsmath,amssymb}
\usepackage{cite}
\usepackage{url}
\usepackage{lscape}
\usepackage{multirow}
\usepackage{bm} 
\usepackage{color}
\let\normalcolor\relax
\newcommand{\be}{\begin{equation}}
\newcommand{\ee}{\end{equation}}
\newcommand{\bes}{\begin{equation*}}
\newcommand{\ees}{\end{equation*}}
\newcommand{\bea}{\begin{eqnarray}}
\newcommand{\eea}{\end{eqnarray}}
\newcommand{\beas}{\begin{eqnarray*}}
\newcommand{\eeas}{\end{eqnarray*}}
\newcommand{\Tr}{\text{Tr}}
\newcommand{\la}{\lambda}

\allowdisplaybreaks

\title{Implications of Unitarity and Charge Breaking Minima in Left-Right Symmetric Model}

\author{Tanmoy Mondal\\
Physical Research Laboratory, Ahmedabad, Gujarat - 380 009, India \\
Indian Institute of Technology, Gandhinagar - 382 424, India\\
E-mail: \email{tanmoym@prl.res.in}
}

\author{Ujjal Kumar Dey\\
Physical Research Laboratory, Ahmedabad, Gujarat - 380 009, India \\
E-mail: \email{ujjaldey@prl.res.in}
}

\author{Partha Konar\\
Physical Research Laboratory, Ahmedabad, Gujarat - 380 009, India \\
E-mail: \email{konar@prl.res.in}
}

\abstract{
We examine the usefulness of the unitarity 
conditions in Left-Right symmetric model which can translate into giving a stronger constraint on 
the model parameters together with the criteria derived from vacuum stability and perturbativity. In this light,
we demonstrate the bounds on the masses of the physical scalars present in the model and 
find the scenario where multiple scalar modes are in the reach of Large Hadron Collider.
We also analyse the additional conditions that can come from charge breaking minima in this context.
}


\begin{document}
\section{Introduction}
\label{sec:intro}
The Large Hadron Collider (LHC) has seen some early success  in discovering the last missing piece of the Standard Model (SM) of the particle physics, 
the Higgs boson~\cite{:2012gu,:2012gk}. 
Nonetheless, there is no substantial evidence yet from physics beyond the Standard Model (BSM) at the LHC, which essentially pushes the 
scale of new physics to higher values. On the other hand, it is widely acknowledged that the SM is a low energy effective theory~\cite{Quigg:2009vq} 
which falls short to explain several theoretical as well as experimental conundrum, such as,  neutrino mass generation, presence of viable dark matter candidate etc.

Left-Right symmetric models (LRSM)~\cite{Pati:1974yy_LR,Mohapatra:1974hk_LR,Mohapatra:1974gc_LR,Senjanovic:1975rk_LR} and its extensions
are very appealing as BSM scenarios. 
These models are advocated for their capability to address the origin of Parity violation in the weak interactions from the
spontaneous breaking of Parity which occurs at the higher energies beyond which Parity is an exact symmetry~\cite{Senjanovic:1975rk_LR}. 
LRSMs also predict the presence of heavy right handed neutrinos explaining the generation of minuscule light neutrino masses
by virtue of the seesaw mechanism~\cite{Mohapatra:1979ia,Mohapatra:1980yp}.  Remarkably, it is also possible 
to realise gauge coupling unification in the non-supersymmetric GUTs where LRSMs appear as low energy effective 
theory~\cite{Arbelaez:2013nga}. 
LRSMs possess extra scalar fields together with the SM Higgs and rather complicated scalar potential emerges consisting of many quartic couplings. 
While some of these couplings can be related directly to the heavy scalar (neutral or charged) masses, 
other couplings contribute in generating the mass splitting among them. 
These quartic couplings can be constrained by imposing theoretical conditions from vacuum stability, perturbativity, as well as, from the unitarity
of scattering amplitudes of longitudinal gauge boson modes. 
Vacuum stability typically restricts the quartic couplings from below, indicating the lower limit, 
whereas perturbativity and unitarity constrain them from above. 
 Unitarity constraint was first analysed by Lee, Quigg and Thacker (LQT)~\cite{Lee:1977eg} 
 for the SM, where they have examined two-body scattering amplitudes involving Higgs boson and 
also longitudinal gauge bosons ($V_L$). Since we are interested in the high energy behaviour of the scattering amplitudes,
it is possible to use unphysical scalars instead of $V_L$s owing to the famous equivalence theorem\footnote{For a pedagogical 
introduction of equivalence theorem, see~\cite{Horejsi:1995jj}.}.

To construct a successfully broken electroweak theory at the low energy, one requires to ensure that the SM-like vacuum is indeed the 
lowest one. To put it another way, if for certain combination of quartic couplings the potential has a minimum where the charged fields 
acquire vacuum expectation values then that coupling combination should be restricted. Thus the analysis of charge breaking (CB) minima 
in principle can put constraints on the scalar quartic couplings. It was first introduced by 
Frere $et.al$~\cite{Frere:1983ag} for supersymmetric theories. In the case of two Higgs doublet model, it has been shown~\cite{Ferreira:2004yd}
that the global minima from the tree-level potential is always charge and $CP$ conserving 
and thus the question of further tunneling into a deeper undesirable minima does not arise here.  One can analyse the aspect of possible CB minima and find the constraints by restricting them in a given model using the formalism described in~\cite{Velhinho:1994np,Ferreira:2004yd}.

Owing to its extended scalar sector as well as the right handed gauge sector, the LRSM offers some interesting 
phenomenological signatures and they were studied in the context of Large Hadron Collider. One of the widely 
studied signal of this model is the so called `golden channel' with characteristic same sign dilepton (SSDL) production along with two associate jets
through $W_R$ and right handed neutrino productions. 
Production of heavy doubly charged scalar can also produce SSDL signals.
There are several other manifestations of LR symmetry that can show up in colliders as signatures of 
TeV scale seesaw, lepton flavour violating processes etc. 
For some of the recent studies regarding LRSM model at the LHC see \cite{ 
Maiezza:2010ic, Nemevsek:2011hz, Frank:2011rb, Chakrabortty:2012pp, Das:2012ii, Dev:2013oxa, Bambhaniya:2013wza, Dutta:2014dba, Bambhaniya:2014cia, 
Bambhaniya:2015wna}. 
Undoubtedly, theoretical bounds can have the ability to restrict the model parameters and thus it affects such extensive 
phenomenological analysis. This is the primary motivation for the study of the theoretical bounds more precisely.
Using the Run-I LHC data at center of mass energy $\sqrt{s} = $ 7 and 8 TeV, the CMS and ATLAS collaborations 
have already set some strong bound on different particles of LR symmetric model. We summarise these bounds in 
table~\ref{my-label}. 
\begin{table}
\begin{center}
\label{my-label}
\begin{tabular}{|c|c|c|}
\hline 
\begin{tabular}[c]{@{}c@{}}Charged heavy gauge boson\\ (TeV)\end{tabular} & \begin{tabular}[c]{@{}c@{}}Doubly charged scalar\\  (GeV)\end{tabular} & \begin{tabular}[c]{@{}c@{}}Heavy neutrino\\ (GeV)\end{tabular} \\ \hline \hline
2.8\cite{CMS:2012uwa,ATLAS-CONF-2013-017}                &  445 \cite{CMS:2012kua} (409~\cite{ATLAS:2012hi})   & 708 \cite{CMS:2012zv,Aad:2012dm}  
\\ \hline \hline
\end{tabular}
\caption{Bounds on different mass scales in Left-Right Symmetric Model from $\sqrt{s} = $ 7 and 8 TeV Run-I LHC data.}
\end{center}
\end{table}

In this article we have considered LRSM constructed with bi-doublet and triplet scalars (LRT)
~\cite{Pati:1974yy_LR, Mohapatra:1974hk_LR, Mohapatra:1974gc_LR,Senjanovic:1975rk_LR}.
Constraints on the scalar sector of this model from vacuum stability 
and perturbativity was discussed in~\cite{Chakrabortty:2013zja,Chakrabortty:2013mha}. In this work we would like to further constrain these quartic couplings of this model by 
imposing unitarity and also demanding that the CB minima is not the global one. 
We find that unitarity is by far the most stringent constraint on the upper limit of the quartic couplings. This, in turn,
 sets the upper limits of masses for physical scalar states, and on the other hand vacuum stability 
restricts masses from below. Using both of these constraints we have analysed and confined the physical scalar masses for this model.

This paper is organised as follows. 
In section~\ref{sec:model},  we briefly describe the model emphasising the extended 
scalar sector  and necessary mass relations with the  quartic couplings. 
We thereafter discuss the calculation in section~\ref{sec:unitarity}, and categorise the effect of unitarity in this model. 
Constraints on couplings as a function of LR symmetry breaking scale are also discussed here. 
Physical scalar masses being more relevant parameters in the search for BSM, we look into the effect of such 
constraints on them. Allowed mass ranges under unitarity constraints together with other restrictions coming from 
vacuum stability and perturbativity are demonstrated in section~\ref{sec:scalmass}. 
In a further investigation we explore the effect of charge breaking minima in this model. We demonstrate the methodology
and final set of conditions in section~\ref{sec:CB}. Additional exhaustive details of the results 
are further supplemented at  the appendix. 
Finally, in section~\ref{sec:conclusion} we summarise and conclude.

\section{Left-Right Symmetric Standard Model with Triplet Scalars}
\label{sec:model}
The Left-Right symmetric models are gauge extension of the SM where an extra $SU(2)_R$ gauge group has been augmented to the 
SM gauge group to incorporate Left-Right symmetry. The full gauge group is $SU(3)\otimes SU(2)_L\otimes SU(2)_R\otimes U(1)_{B-L}$.
After spontaneous breaking, $SU(2)_R\otimes U(1)_{B-L}$ will break to $U(1)_Y$ giving rise to SM gauge group.  
The scalar sector contains one left handed triplet ($\Delta_L$), one right handed triplet ($\Delta_R$) along with a bi-doublet ($\Phi$),
which, in component form, can be written as:
\begin{center}
 \be\label{eq:scalar-structure}
\hskip 25pt\Delta_{L,R}   = \left( \begin{array}{cc}
         \delta_{L,R}^+/\sqrt{2} & \delta_{L,R}^{++} \\      
         \delta_{L,R}^0 & -\delta_{L,R}^+/\sqrt{2}
        \end{array} 
        \right)
  \hskip 10pt , \hskip 10pt       
 \Phi =\left( \begin{array}{lr}
         \phi_1^0 \;&\; \phi_1^+\\ 
         \phi_2^- & \phi_2^0
        \end{array} 
        \right).
\ee
\end{center}
Vacuum expectation values for these fields are given by:
\be\label{eq:LRT_vev_struct} 
\left< \Phi \right>  =\left( \begin{array}{cc}
                        v_1 & 0\\
                        0      \;&\; v_2
                       \end{array}\right) , \hskip 20 pt                       
\left< \Delta_L \right>  =\left( \begin{array}{lr}
                        0   & 0\\
                        v_L\; & \;0
                       \end{array}\right) , \hskip 20 pt                       
\left< \Delta_R \right>  = \left(\begin{array}{lr}
                        0   & 0\\
                        v_R \;&\; 0
                       \end{array} \right).
\ee
The Left-Right symmetry is broken to the SM at the $v_R$ scale. 
Breaking of electroweak symmetry  to $U(1)_{\rm em}$ was triggered 
by\footnote{It is worth mentioning that the vev of $\Delta_L(\langle\delta_L^0\rangle=v_L)$ 
affects tree level $\rho$ parameter~\cite{Agashe:2014kda} and hence it is constrained to be 
$v_L \leq 2.5$ GeV.} $v_1,v_2$ and $v_L$. The most general form of LRT scalar potential is 
discussed extensively in \cite{Mohapatra:1980yp,PhysRevD.39.870,Deshpande:1990ip} and for our analysis 
we use the form of  the potential given in \cite{Deshpande:1990ip}. The potential
is written in appendix~\ref{app:lrt-pot}, containts fifteen quartic couplings. 
The minimization condition of the potential yields the vev-seesaw relation {\it i.e.}, 
\be
\la_{14}v_1^2+\la_{13}v_1 v_2 + \la_{15}v_2^2 = (2\la_5-\la_7) v_L v_R~,
\ee
which must be satisfied for successful breaking of the LR symmetry. 
It was discussed~\cite{Deshpande:1990ip} in detail 
how the three quartic couplings viz. $\la_{13},\la_{14}\textrm{ and }\la_{15}$ can be set to zero along with the vev of $\Delta_L$.
It is interesting to note that 
beside satisfying the vev-seesaw relation, this choice of parameters 
comes with additional benefit of considerably reduced degree of fine tuning in the model 
as compared to one with non-zero parameters~\cite{Dekens:2014ina}. 
In addition, we consider that $v_2=0$. As far as the physical scalars are concerned, this model contains 
twenty real scalar fields which finally give rise to two doubly charged, two singly charged, 
four neutral $CP$ even and two neutral $CP$ odd massive scalars. Lightest of these neutral $CP$ even scalar ($H_0^0$), 
is assigned as Standard Model Higgs. We set this mass at $125$ GeV for our present calculation.
However, all other heavy scalars are associated with a much heavier scale $v_R$ and squeezed to that 
mass scale. The leading order terms for scalar masses~\cite{Duka:1999uc,Czakon:1999ue_WR_mudecay}  are,
\begin{subequations}\label{eq:LR-Masses}
\bea 
M_{H_0^0}^2 &\simeq& 2 \, \la_1 \,v_1^2 ,\\
M_{H_1^0}^2 \simeq  M_{A_1^0}^2 \simeq M_{H_2^\pm}^2 &\simeq& \frac{1}{2} \la_{12}\, v_R^2,  \\
M_{H_2^0}^2 &\simeq& 2 \, \la_5 \, v_R^2, \\
M_{H_3^0}^2 \simeq  M_{A_2^0}^2 \simeq M_{H_1^\pm}^2 \simeq M_{H_1^{\pm\pm}}^2 &\simeq& \frac{1}{2} (\la_7 - 2\la_5) \, v_R^2, \\
M_{H_2^{\pm\pm}}^2 &\simeq& 2\, \la_6 \, v_R^2.
\eea
\end{subequations}

Note that, in the expression of heavy scalar masses, quartic couplings  $\lambda_1, \lambda_5, \lambda_6, \lambda_7,$ and $\lambda_{12}$ 
comes with the common heavy state mass scale $v_R$. All other quartic couplings contribute in sub-leading 
effect to these heavier masses with a factor proportional to the EW vev, $v_1$.

\section{Unitarity Constraints}
\label{sec:unitarity}

Any scattering amplitude can be written as an infinite sum of partial waves, in the form,
\begin{align}
\mathcal{M}(\theta) = 16\pi\sum_{l=0}^{\infty}a_l\,(2l+1)\,P_l(\cos\theta),
\end{align}
where $a_{l}$ is the scattering amplitude of order $l$, $\theta$ is the scattering angle and 
$P_{l}(\cos\theta)$ is $l^{\rm th}$-order Legendre polynomial. In SM, by analysing the two-body 
scattering between longitudinal gauge bosons and Higgs it was shown in the seminal paper by LQT~\cite{Lee:1977eg} 
that unitarity of $S$-matrix constrains the zeroth partial wave amplitude as, $|a_{0}|\leq 1$ which in turn 
restricts the Higgs quartic coupling and therefore constrains the Higgs mass from the above. Eventually the 
unitarity constraint can be recast as,
\begin{align}
\label{eq:ampunitary}
 |\mathcal{M}| \leq 8\pi, 
\end{align}
where $\mathcal{M}$ is the full tree level 
matrix element. This method can be extended to the scenario where extra scalar fields are 
present~\cite{Horejsi:2005da, Kanemura:1993hm, Akeroyd:2000wc, Das:2014fea}. Thus in the present scenario, 
we also consider the appropriate two-body channels. By virtue of equivalence theorem, in the high energy 
limit, one can use the unphysical scalars instead of original longitudinal components of the gauge bosons. Thus 
the relevant $2\to 2$ scatterings will get contributions from the quartic couplings; the contribution from 
trilinear couplings can safely be ignored due to the fact that the diagrams resulting from the trilinear 
couplings will have an $E^{2}$-suppression coming from the intermediate propagators. So we need to find the 
matrix elements for relevant $2\to 2$ processes. Accordingly an $S$-matrix can be constructed by taking different 
two-particle states as rows and columns and each entry of that matrix will give the scattering amplitude between 
the corresponding 2-particle state in the row and the 2-particle state in the column. Clearly, the unitarity constraints 
(eq.~\ref{eq:ampunitary}) manifest themselves as bounds on the eigenvalues of this matrix.  

\begin{table} [t]
\centering
\begin{tabular}{|c|c|c|c|c|c|}
\hline
Charge($q$)                                                            & 0                              & 1        & 2                     & 3    & 4                  \\ \hline \hline
\begin{tabular}[c]{@{}c@{}}Number of states\\  (general)\end{tabular}  & $\frac{n(n+1)}{2}+s^{2}+d^{2}$ & $s(n+d)$ & $nd+\frac{s(s+1)}{2}$ & $~~sd~~$ & $\frac{d(d+1)}{2}$ \\ \hline
\begin{tabular}[c]{@{}c@{}}Number of states\\ (LRT Model)\end{tabular} & 56                             & 40       & 26                    & 8    & 3                  \\ \hline
Independent constraints                                                & 29                             & 14       & 16                    & 3    &  2                  \\ \hline  
\end{tabular}
\caption{Number of all possible $q$-charged 2-particle states constructed from $n$ neutral, $s$ singly charged and $d$ doubly 
charged fields. The second row gives a generic result whereas the third row is specific to LRT model. In the fourth row we mention 
the number of independent eigenvalues arising from each $q$-charged $S$-matrix. }
\label{tab:2-particle-states}
\end{table}

In our case, the 2-particle states are made of the component fields corresponding to the parametrization of 
eq.~\ref{eq:scalar-structure}. As evident from eq.~\ref{eq:scalar-structure}, the model contains neutral, 
singly charged and doubly charged states. Using them we constructed all possible $q$-charged 2-particle 
states, where $q$ can be anything from zero to four. If one has $(n)$-neutral, $(s)$-singly charged and $(d)$-doubly 
charged fields then the number of all possible 2-particle states are tabulated in the second row of table~\ref{tab:2-particle-states}.

In this present case of Left-Right symmetric model the values of $n, s$ and $d$ are 8, 4 and 2 respectively. Hence, 
we computed all the respective $q$-charged states  (as listed in the third row of table~\ref{tab:2-particle-states}) 
 and composed the corresponding $S$-matrix. Finally, we calculate the eigenvalues of these matrices and restrict them to 
have an upper limit of $8\pi$ (cf.~\cite{Horejsi:2005da}) to derive the constraint equations.
One can find that some of the eigenvalues repeat itself and hence the independent number 
of unitarity constraints are fewer compared to that of total number of eigenvalues which has been tabulated 
in the fourth row of same table. 
Complete list of expressions of all the unitarity constraints and eigenvalues are provided and discussed in the appendix~\ref{app:suppl}.

 To illustrate the effect of unitarity, we follow the similar simplified prescription as in~\cite{Chakrabortty:2013zja} and 
consider the quartic couplings\footnote{To obey the charge breaking conditions, discussed in section~\protect \ref{sec:CB}, 
we chose vanishing $\la_{11}$ and it remains so even during the RG evolution.} $\la_2, \la_3, \la_4, \la_8, \la_9, \la_{10}$ (=$\la_{u}$) 
universal as they only contribute in mass splittings between the heavy scalar states. 
Since these couplings are not accessible at the collider, they can only be constrained by using vacuum stability, 
perturbativity and unitarity. While the effect of vacuum stability and perturbativity was extensively 
discussed in~\cite{Chakrabortty:2013zja,Chakrabortty:2013mha},
here we would like to analyse the bound from unitarity of the $S$-matrix and demonstrate combined results together. 

\begin{figure} [t]
 \centering
 \includegraphics[width=7cm,angle=-90]{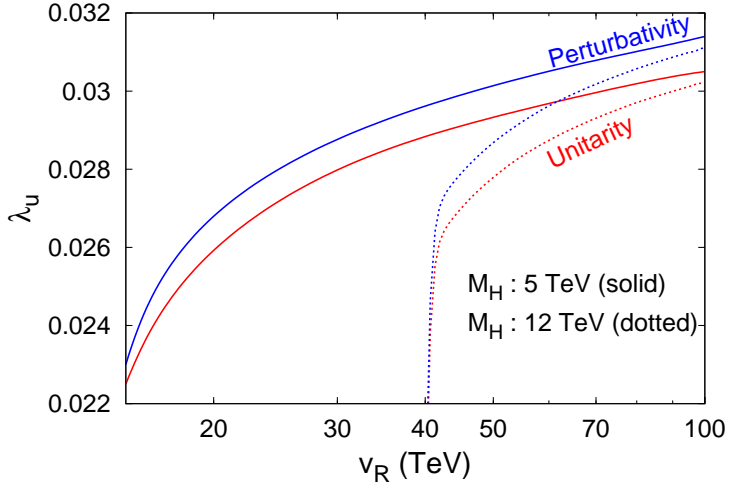}
 \caption{Constraints on the universal quartic coupling $\lambda_u$ for LR model coming from unitarity (red) and perturbativity (blue) bounds for multi-TeV region of Left-Right symmetry breaking scale $v_R$. Also, $v_R$ scale is heavy enough ($> 10 $ TeV) to satisfy the constraints.  Stringent bounds are coming from unitarity. Two different sets of heavy scalar states ($M_H$), viz., 5 TeV and 12 TeV are considered for demonstration. Charge breaking conditions discussed in section~\protect \ref{sec:CB} are also implied here.
}
 \label{fig:unitarity}
\end{figure}

Using the renormalisation group evolution equations~\cite{Rothstein:1990qx_LRT}, we evaluate the quartic couplings at each scale 
all the way up to the Planck scale ($M_{Pl}$) and impose constraints coming from unitarity, vacuum stability and perturbativity.
We extract maximum allowed values for these couplings 
keeping all heavy scalar masses degenerate to some high scale ($M_H$).  
Low energy data like $K_L - K_S$ mass difference restricts $W_R > 3.5$ 
TeV~\cite{Beall:1981ze_kLkS,Langacker:1989xa_WR,Czakon:2002wm_mudecay,Chakrabortty:2012pp_WR_mudecay} and LHC direct search 
limit is $W_R >  3$ TeV~\cite{ATLAS:2014wra,Khachatryan:2014tva}. One can easily translate these bounds to the LR symmetry breaking 
scale $v_R$ using the relation: 
\be
M_{W_R^\pm}^2 = \frac{1}{4}g_{_{2}}^2\,\left(v_1^2\,+\,v_R^2\right).
\ee
We have adopted $v_R$ to be heavy enough ($>10$ TeV) for our analysis so that these bounds are easily satisfied.

Figure~\ref{fig:unitarity} demonstrates both the constraints  coming from unitarity (red curves), as well as perturbativity (blue curves)
 on the universal quartic coupling $\lambda_u$ for Left-Right symmetric model. Multi-TeV region of Left-Right symmetry breaking scale $v_R$ is considered. 
 Also, two different sets of heavy scalar states ($M_H$), assuming heavy scalar states are nearly degenerate, are considered for presentation.
Clearly, for a particular value of Left-Right symmetry breaking scale ($v_R$) unitarity bounds put severe constraints on quartic couplings 
compared to that of coming from perturbativity bounds.  
We have implemented the perturbativity bound\footnote{In~\cite{Chakrabortty:2013zja} this bound was 
set to a more conservative value of unity.} as  $|\la_i| < 4\pi,\;\forall \; i \in [0,15]$.

\section{Constraints on Physical Scalar Masses}
\label{sec:scalmass}

In the previous section we have demonstrated the usefulness of unitarity to constrain the quartic coupling in the Left-Right symmetric model.
Now, we turn to look some more phenomenologically useful aspects, in an era, when the Large Hadron Collider (LHC) is expected to explore new physics at multi-TeV scale.
Here we use vacuum stability along with unitarity and perturbativity to constrain the physical scalar mass states. 
Vacuum stability criteria for this model are calculated using copositivity of symmetric matrices in~\cite{Chakrabortty:2013mha} and combined conditions read as:
\be\label{eq:vs}
\la_1 \geq 0 \;\;\;{\bm\land}\;\;\; \la_5 \geq 0 \;\;\;{\bm\land}\;\;\; \la_5+\la_6 \geq 0 \;\;\;{\bm\land}\;\;\;  16\;\la_1\,\la_5 - \la_{12}^2 \geq 0.
\ee

\begin{figure}
 \centering
 \includegraphics[width=7cm,angle=-90]{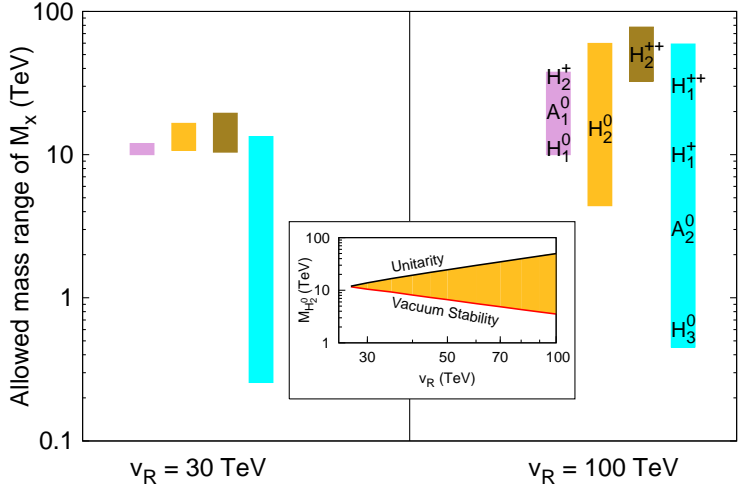}
 \caption{Allowed mass range for four sets of heavy scalar states ($M_X$) after imposing all constraints coming 
from vacuum stability, unitarity, as well as perturbativity at each scale all the way up to Planck scale ($M_{Pl}$).
Two different sets of Left-Right symmetry breaking scale $v_R$  is considered, which are 30 TeV and 100 TeV.
The bound $M_{H_1^0} > 10$ TeV has also been taken into account. Here $\la_{u}$ is set to the value 0.01
which is much below the unitarity bound and this is evident from the figure~\protect \ref{fig:unitarity}.
Inset shows how one set of heavy scalar mass (e.g., $M_{H^0_2}$) is constrained from vacuum stability (red)  
and unitarity (black) bounds  over a continuous range of  $v_R$. 
}
 \label{fig:mass}
\end{figure}
To explore the allowed mass range of physical scalars, at LR symmetry breaking scale ({\it i.e.}, $v_R$ scale) 
we randomly vary  the quartic couplings\footnote{The parameter $(\la_7 - 2\la_5)$ sets mass scale for some scalars 
and instead of $\la_7$ we randomly vary the difference $(\la_7 - 2\la_5)$ in the range [$0,4\pi$] 
to ensure that no unphysical mass scale appears in the model.} $\la_5,\la_6,\la_7$ and $\la_{12}$ 
in their allowed range\footnote{In general quartic coupling can take any value from $[-4\pi,4\pi]$ 
but here these couplings can not be negative as it will lead to tachyonic states.} $[0,4\pi]$
and estimate the corresponding mass scales.
These quartic couplings run according to their respective RGEs~\cite{Rothstein:1990qx_LRT} and we ensure that 
the quartic couplings obey all the conditions coming 
from vacuum stability, unitarity, as well as perturbativity at each scale below $M_{Pl}$. 
The input quartic couplings which obey these conditions, till $M_{Pl}$, are interpreted as the accepted mass scale of 
physical scalars using eq.~\ref{eq:LR-Masses}.  

In figure~\ref{fig:mass} we demonstrate this allowed mass range for four sets of heavy scalar states listed in 
eq.~\ref{eq:LR-Masses} (except first one, which is actually input parameter mass) after imposing all constraints 
as described above. Below we present the 
detailed discussion about each of these sets of heavy scalars. 
This is demonstrated for two different LR symmetry breaking scale viz. 30 TeV and 100 TeV and corresponding
mass ranges are also tabulated in table~\ref{tab:mass}.
We also display, in the inset of figure~\ref{fig:mass}, how one set of heavy scalar mass (e.g., $M_{H^0_2}$) 
is constrained from vacuum stability (red)  and unitarity (black) over a continuous range of  $v_R$.

From these considerations one can make following observations about the allowed mass range:

\begin{itemize}
 \item To suppress the FCNC effects the fields $H_1^0$ and $A_1^0$ should be heavy 
 $\sim 10$ TeV~\cite{Ecker:1983uh,Mohapatra:1983ae,Pospelov:1996fq}. 
 We use this information to limit the corresponding quartic coupling $\la_{12}$ from below at $v_R$ scale and on the 
 other hand perturbativity restricts the coupling from above. This can be seen in the purple bar (left most region) 
 where allowed mass range for $M_{H_1^0}, M_{A_1^0} \textrm{ and } M_{H_2^\pm}$ is very narrow for small $v_R$ value. 
 Large $v_R$  relaxes the perturbativity bound and larger region is allowed. 
This also sets an minimum allowed value of LR breaking scale $v_R$ coming  from vacuum stability and perturbativity, 
which can also be marked from the inset plot. However, this would make sense only
 if FCNC bound is robust. Non-minimal  LR model can avoid FCNC bound and  few TeV scale $H_1^0$ is allowed~\cite{Guadagnoli:2010sd}. 
 \item Allowed range of $M_{H_2^0} (=2 \, \la_5 \, v_R^2 )$ is depicted in orange/yellow band. 
 To explain its behaviour we add an inset in the figure~\ref{fig:mass} where a continuous variation 
 of $M_{H_2^0}$ with $v_R$ is shown. Since $\la_5$ and $\la_{12}$ are coupled through vacuum stability condition 
 (cf. eq.~\ref{eq:vs}) the minimum allowed value of $\la_5$ is fixed at $v_R$ scale which sets the scale of $M_{H_2^0}$. 
 For fixed $M_{H_{1}^{0}}$ (10 TeV), higher value of $v_R$ supports lower $\la_{12}$ which eventually decreases minimum allowed value for $\la_5$. Maximum allowed 
 value of $\la_5$ is restricted from unitarity. As evident from the figure, $M_{H_2^0}$ can be light enough \textit{i.e.,} $\mathcal{O}$ (TeV)
   for higher values of $v_R$. 

 \item Mass of $M_{H_2^{\pm\pm}}$ is defined  by the quartic coupling $\la_6$. With low initial value, 
 this coupling decreases with energy and eventually becomes negative leading to tachyonic states. 
 To get rid of tachyonic states the boundary value for $\la_6$ at $v_R$ scale should be high enough 
 and this leads to relatively higher mass states for $M_{H_2^{\pm\pm}}$ as shown in olive band.  
 \item The parameter $(\la_7 - 2\la_5)$ governs mass scale for\footnote{Note 
 that LEP II data yields a lower bound on the mass of $H_{3}^{0}$, which is about 55 GeV~\cite{Datta:1999nc}.} 
 $M_{H_3^0}, M_{A_2^0}, M_{H_1^\pm}, \textrm{and } M_{H_1^{\pm\pm}}$ and it can become very small as it is not 
 constrained from below via vacuum stability. But the mass scale will shift as there are secondary contributions coming 
 from universal quartic couplings and EW breaking vev $v_1$. The cyan bar represents the allowed range for these scalars. 
 In principle contribution to these scalars coming from LR breaking vev can be zero and in that case the secondary contribution 
 of $\mathcal{O}(100)$ GeV will set the mass scale. The minimum values shown in figure~\ref{fig:mass} are nothing but numerical artifact where the 
 coupling is already very small ($\sim 10^{-5}$).  
\end{itemize}

  
\begin{table}
\begin{center}
\begin{tabular}{|c|c|c|c|c|}
\hline
$v_R$ & $M_{H_1^0}, M_{A_1^0}, M_{H_2^\pm}$  & $M_{H_2^0}$ & $M_{H_2^{\pm\pm}}$& {$M_{H_3^0}, M_{A_2^0}, M_{H_1^\pm}, M_{H_1^{\pm\pm}}$}\\[2ex]
(TeV)&(TeV)&(TeV)&(TeV)&(TeV)\\\hline \hline
30  & {$10-12$}         & {$10.5-16.5$}  & {$10.5-20$}      & {$\mathcal{O}(0.1)-13.5$ }   \\[1.1ex] \hline
100 & {$10-37.5$}        & {$4.4-60$}   & {$33-78$}      & {$\mathcal{O}(0.1)-59$ }     \\[1.1ex] \hline
\end{tabular}
\caption{Allowed mass range in TeV for two different $v_R$ scale. 
These are approximated values as there will be secondary contributions 
which will shift masses upward by $\mathcal{O}(100)$ GeV 
which is insignificant except for the lower limit of the last column.}
\label{tab:mass} 
\end{center}
\end{table}

\section{Charge Breaking Minima of Tree Level Potential}
\label{sec:CB}
It is believed that the present Universe is at the SM ground state where only neutral component of a doublet scalar gets a vev. 
Nevertheless, if a model contains  multiple scalars, existence of charge breaking global minima is also plausible which may lead 
to disastrous results, like non-conservation\footnote{In principle it is also possible that a $CP$ breaking global minima 
can arise, although its not the case here as we have restricted all $CP$ violating phases to zero.}  of electric charge, and massive photons. 
In that case, even if the SM minima is a local one, the catastrophe can possibly be averted 
provided that the tunneling time to a deeper charge breaking minima is larger than the age of the Universe.

Before we begin, let us give a brief description introducing some notations.
We can write the scalar potential in a matrix form as,
\be
 V=A^T X + X^TBX;  
 \ee
where, the column matrix $A$ contains the quadratic part of the potential and the symmetric matrix $B$ carries quartic couplings. 
$X = \{x_1,x_2,\cdots,x_n\}^T$ is a vector which contains the combination of scalar fields. 
Using this potential, it is straightforward to get
\be
V' = A+BX_{SM} \textrm{\; where\; }V'_i = \frac{\partial V}{\partial x_i},
\ee
here, $X$ is replaced by the $X_{SM}$, which is the basis vector evaluated at some SM-like potential minimum. 
Value of the potential at that minima is given by,
\be\label{eq:pot-value}
V_{SM}=\frac{1}{2}A^T\,X_{SM} = -\frac{1}{2} X_{SM}^T\,B\,X_{SM}.
\ee
Now, if there exist another minima of the potential $V$, we  need to find out which 
minima is the global one. Moreover, global minima should not break charge or color quantum number. 
Let us consider that $X_{CB}$ be basis vector at another minima which breaks electric charge. The minimization condition 
ensures:
\be
\frac{\partial V}{\partial X}\bigg\vert_{X=X_{CB}}\hskip -10pt=0 \hskip 15pt \Rightarrow  \hskip 15pt  A+B\,X_{CB} = 0.
\ee
The potential at that charge breaking minima, $V_{CB}$ can be written in a similar fashion of eq.~\ref{eq:pot-value} by replacing 
the vector $X_{SM}$ by the new vector $X_{CB}$. Using the above equations one can easily show that the difference 
of potential between the charge breaking and the SM-like minima as, 
\be\label{eq:vcbmvn}
V_{CB}-V_{SM} = \frac{1}{2} X_{CB}^T V'.
\ee
Clearly, the charge breaking minima is not a global one if $(V_{CB}-V_{SM}) > 0$. 
Accoutered with this general discussion, we are now in a position to compute the same in the Left-Right symmetric model with triplet scalars. 
The basis vector $X$ in this model can be written as:
 \be
 X=\begin{pmatrix}
 \Tr\big[\Phi^\dagger \Phi\big] \\
 \Tr\big[\tilde{\Phi}\Phi^\dagger\big]+\Tr\big[\tilde{\Phi}^\dagger \Phi\big]\\
 \Tr\big[\tilde{\Phi}\Phi^\dagger\big]-\Tr\big[\tilde{\Phi}^\dagger \Phi\big]\\
 \Tr\big[\Delta_R^\dagger \Delta_R\big]\\
 \Tr\big[\Delta_R \Delta_R\big]+\Tr\big[\Delta_R^\dagger \Delta_R^\dagger\big]\\
  \Tr\big[\Delta_R \Delta_R\big]-\Tr\big[\Delta_R^\dagger \Delta_R^\dagger\big]\\
   \Tr\big[\Delta_L^\dagger \Delta_L\big]\\
 \Tr\big[\Delta_L \Delta_L\big]+\Tr\big[\Delta_L^\dagger \Delta_L^\dagger\big]\\
  \Tr\big[\Delta_L \Delta_L\big]-\Tr\big[\Delta_L^\dagger \Delta_L^\dagger\big]\\
\{\Tr\big[\Phi\Delta_R\tilde{\Phi}^\dagger\Delta_L^\dagger\big]+ \Tr\big[\Phi^\dagger\Delta_L\tilde{\Phi}\Delta_R^\dagger\big]\}^{1/2} \end{pmatrix}.
 \ee
From the explicit form of the potential, as provided in eq.~\ref{eq:vlrt}, it is evident that $A$ is nothing but a column matrix comprised of the quadratic coefficients $\mu_{i}~(i=1,2,3)$ and is given by, 
\be 
A=\{-\mu_1^2,-\mu_2^2,0,-\mu_3^2,0,0,-\mu_3^2,0,0,0,0\}^T.
\ee
The matrix $B$ is a symmetric matrix and is written explicitly in appendix~\ref{app:bmatrix}. For SM-like minima the 
vev structure is given in eq.~\ref{eq:LRT_vev_struct} and we have computed $V'$ at this minima as
\be\label{eq:vprime}
V'=\left\lbrace0, 0, -iv_R^2 \la_{11}, 0, 0, 0, \frac{1}{2}(v_R^2\la_7+v_1^2\la_9), 0, 0, 0\right\rbrace ^T.
\ee

As we have already mentioned that for any charge breaking minima, one simply needs to ensure
that the difference $(V_{CB}-V_{SM})$ is positive to confirm that the SM-like minima 
is the global one. 
For illustration, if the charged scalar field $\delta_L^{++}$ of eq.~\ref{eq:scalar-structure} 
gets a vev $v_C$, then new field vector at the CB minima is 
\be
X_{CB} = \left\lbrace\frac{v_1^2}{2},\, 0,\, 0,\, \frac{v_R^2}{2},\, 0,\, 0,\, v_C^2,\, 0,\, 0,\, \frac{v_1\,v_C}{\sqrt{2}} 
 \right\rbrace ^T,
\ee
which produces the differences between {\it normal} and {\it charge breaking} minima
\be
V_{CB}-V_{SM} = \frac{1}{4}\,v_C^2(\la_7\,v_R^2+\la_9\,v_1^2).
\ee
Since $v_R \gg v_1$, charge breaking global minima is only possible if $\la_7$ is large negative.
Simple assumptions like all the quartic couplings are real and positive can safeguard from any such problems and we chose so for our analysis.

In principle, it is possible that more than one field directions get vev in a CB minima. 
Since there are six different charged field directions, number of $n$-field CB minima is $\binom{6}{n}$. Aforementioned 
example of $\delta_L^{++}$ getting vev is 1-field CB minima. 
We need to consider all of these directions to ensure that the SM is the lowest minima. 
From eq.~\ref{eq:vcbmvn} it is evident that if we can ensure that elements of the column matrix $V'$ in eq.~\ref{eq:vprime}
are positive then for any CB field directions the $(V_{CB}-V_{SM})$ will remain positive independent of the form of the 
CB minimum, $X_{CB}$. Hence the final set of conditions are, 
\be\label{eq:cbconditions}
 \la_7 > - \left(\frac{v_1^2}{v_R^2}\right)\,\la_9.
\ee

Since we have assumed that all the quartic couplings are positive, CB condition is readily satisfied. Also note that eq.~\ref{eq:vprime} also implies a condition on $\la_{11}$ which can be taken care of by setting it equal to zero.

One can recognise that, for some of the CB directions, it is possible that CB and SM minima become degenerate {\it i.e.}, $(V_{CB}-V_{SM})=0$. As we have already mentioned that the presence of charge breaking global minima will break the electric charge conservation and if normal and charge breaking global minima has to coexist then the present Universe must choose the normal minima out of all the possibilities. After choosing the normal minima its not possible to tunnel to degenerate charge breaking minima as it would require an infinite amount of energy. However, it is worth mentioning that the radiative corrections may lift this degeneracy,  which is beyond the scope of present study.

\section{Conclusion}
\label{sec:conclusion}
Being a very simple gauge group extension of the SM and giving a rich dividend in BSM phenomena, 
Left-Right symmetric models are phenomenologically interesting in their own right. 
The scalar sector of this model is quite rich due to the fact that an enlarged scalar sector is 
required to get a mechanism for breaking the Left-Right symmetric group to SM gauge group. In the present work we 
analysed the scalar sector of the Left-Right symmetric standard model with triplet scalars in the light of various 
theoretical and experimental constraints.

The scalar sector comprised of one bi-doublet, one left handed and one right handed triplet scalar ultimately give 
rise to fourteen physical scalars. Lightest among them is expected to  
be the recently discovered Higgs boson with mass around
125 GeV. We constrain the masses of the other physical scalars by using the unitarity constraints. We obtain these 
constraints by evaluating the zeroth order partial wave amplitude of various $2\to 2$ scatterings. 
We find that for any Left-Right symmetry breaking scale, unitarity bounds put severe constraints on quartic couplings 
compared to that of coming from perturbativity.
We also demonstrated that 
some of the physical scalars can have the mass in the TeV range which can have interesting LHC prospects. 
It is to be noted that the masses of these scalars are dependent on the Left-Right symmetry breaking scale $v_{R}$ 
and consequently obtained bounds are highly sensitive to this $v_{R}$. 

We also discussed the charge breaking minima of the tree-level potential. 
We derive the conditions that the quartic couplings must satisfy to avoid the charge breaking minima. 
These conditions were implemented to restrict physical scalar masses and the quartic couplings.

\bigskip
\acknowledgments
This work was funded by Physical Research Laboratory (PRL), Department of Space (DoS), India. 
Authors want to thank Goran Senjanovic, N.G. Deshpande and Fred Olness for useful discussions. 


\appendix

\section{Scalar Potential}
\label{app:lrt-pot}
Full scalar potential for LRT model can be written as:
\begin{align}
\label{eq:vlrt}
V_{LRT}(\Phi,& \Delta_L,\Delta_R) = \notag \\
    &- \mu_1^2\bigg\{\Tr\big[\Phi^\dagger \Phi\big]\bigg\} 
    - \mu_2^2\bigg\{\Tr\big[\tilde{\Phi}\Phi^\dagger\big]+\Tr\big[\tilde{\Phi}^\dagger \Phi\big]\bigg\} 
    - \mu_3^2\bigg\{ \Tr\big[\Delta_L^\dagger \Delta_L\big]+\Tr\big[\Delta_R^\dagger \Delta_R\big] \bigg\}  \notag \\
    &+ \lambda_1\bigg\{\Big(\Tr\big[\Phi^\dagger \Phi\big]\Big)^2\bigg\} +
    \lambda_2\bigg\{ \Big(\Tr\big[\tilde{\Phi}\Phi^\dagger\big]\Big)^2+\Big(\Tr\big[\tilde{\Phi}^\dagger \Phi\big]\Big)^2 \bigg\} 
    + \lambda_3\bigg\{\Tr\big[\tilde{\Phi}\Phi^\dagger\big]\Tr\big[\tilde{\Phi}^\dagger \Phi\big] \bigg\}\notag \\
    &+ \lambda_4 \bigg\{ \Tr\big[\Phi^\dagger \Phi\big]\Big(\Tr\big[\tilde{\Phi}\Phi^\dagger\big]
    +\Tr\big[\tilde{\Phi}^\dagger \Phi\big]\Big) \bigg\}
    +\lambda_5\bigg\{ \Big(\Tr\big[\Delta_L \Delta_L^\dagger\big]\Big)^2+\Big(\Delta_R \Delta_R^\dagger\Big)^2 \bigg\} \notag \\
    &+ \lambda_6 \bigg\{\Tr\big[\Delta_L \Delta_L\big]\;\Tr\big[\Delta_L^\dagger \Delta_L^\dagger\big]
    +\Tr\big[\Delta_R \Delta_R\big]\;\Tr\big[\Delta_R^\dagger \Delta_R^\dagger\big]  \bigg\}
    + \lambda_7 \bigg\{\Tr\big[\Delta_L\Delta_L^\dagger\big]\;\Tr\big[\Delta_R\Delta_R^\dagger\big]\bigg\} \notag \\
    &+ \lambda_8[\Delta_L\Delta_L^\dagger\big] \bigg\{\Tr\big[\Delta_L\Delta_L^\dagger\big]\;
    \Tr\big[\Delta_R\Delta_R^\dagger\big] \bigg\}
    + \lambda_9 \bigg\{\Tr\big[\Phi^\dagger \Phi\big]\Big(\Tr\big[\Delta_L\Delta_L^\dagger\big]
    +\Tr\big[\Delta_R\Delta_R^\dagger\big]\Big)\bigg\} \notag \\
    &+ (\lambda_{10}+i\,\lambda_{11}) 
    \bigg\{\Tr\big[\Phi\tilde{\Phi}^\dagger\big]\Tr\big[\Delta_R\Delta_R^\dagger\big] 
    + \Tr\big[\Phi^\dagger\tilde{\Phi}\big]\Tr\big[\Delta_L\Delta_L^\dagger\big]\bigg\} \notag \\
    &+ (\lambda_{10}-i\,\lambda_{11}) 
    \bigg\{\Tr\big[\Phi^\dagger\tilde{\Phi}\big]\Tr\big[\Delta_R\Delta_R^\dagger\big] 
    + \Tr\big[\tilde{\Phi}^\dagger\Phi\big]\Tr\big[\Delta_L\Delta_L^\dagger\big]\bigg\} \notag \\
    &+ \lambda_{12} \bigg\{ \Tr\big[\Phi \Phi^\dagger \Delta_L \Delta_L^\dagger\big]
    +\Tr\big[\Phi^\dagger \Phi \Delta_R \Delta_R^\dagger\big]\bigg\}
    + \lambda_{13} \bigg\{\Tr\big[\Phi\Delta_R\Phi^\dagger\Delta_L^\dagger\big]
    +\Tr\big[\Phi^\dagger\Delta_L\Phi\Delta_R^\dagger\big] \bigg\} \notag \\
    &+ \lambda_{14}\bigg\{\Tr\big[\tilde{\Phi}\Delta_R\Phi^\dagger\Delta_L^\dagger\big]
    +\Tr\big[\tilde{\Phi}^\dagger\Delta_L\Phi\Delta_R^\dagger\big]\bigg\} 
    + \lambda_{15} \bigg\{\Tr\big[\Phi\Delta_R\tilde{\Phi}^\dagger\Delta_L^\dagger\big]
    +\Tr\big[\Phi^\dagger\Delta_L\tilde{\Phi}\Delta_R^\dagger\big]\bigg\},
\end{align}
where all the coupling constants are real.

\section{The Scalar Quartic Coupling Matrix}\label{app:bmatrix}
\begin{scriptsize}
$$
\left(
\begin{array}{cccccccccc}
 2 \lambda_1(x) & 2 \lambda_4(x) & 0 & \lambda_9(x) & 0 & 0 & \lambda_9(x) & 0 & 0 & 0 \\
 2 \lambda_4(x) & 2 \left(2 \lambda_2(x)+\lambda_3(x)\right) & 0 & \lambda_{10}(x)+\lambda_{11}(x) & 0 & 0 & \lambda
   _{10}(x)+\lambda_{11}(x) & 0 & 0 & 0 \\
 0 & 0 & 2 \left(2 \lambda_2(x)-\lambda_3(x)\right) & \lambda_{11}(x)-\lambda_{10}(x) & 0 & 0 & \lambda_{10}(x)-\lambda_{11}(x) & 0
   & 0 & 0 \\
 \lambda_9(x) & \lambda_{10}(x)+\lambda_{11}(x) & \lambda_{11}(x)-\lambda_{10}(x) & 2 \lambda_5(x) & 0 & 0 & \lambda_7(x) & 0 & 0
   & 0 \\
 0 & 0 & 0 & 0 & 2 \lambda_6(x) & 0 & 0 & 2 \lambda_8(x) & 0 & 0 \\
 0 & 0 & 0 & 0 & 0 & -2 \lambda_6(x) & 0 & 0 & -2 \lambda_8(x) & 0 \\
 \lambda_9(x) & \lambda_{10}(x)+\lambda_{11}(x) & \lambda_{10}(x)-\lambda_{11}(x) & \lambda_7(x) & 0 & 0 & 2 \lambda_5(x) & 0 & 0
   & 0 \\
 0 & 0 & 0 & 0 & 2 \lambda_8(x) & 0 & 0 & 2 \lambda_6(x) & 0 & 0 \\
 0 & 0 & 0 & 0 & 0 & -2 \lambda_8(x) & 0 & 0 & -2 \lambda_6(x) & 0 \\
 0 & 0 & 0 & 0 & 0 & 0 & 0 & 0 & 0 & 2 \lambda_{12}(x) \\
\end{array}
\right)
$$
\end{scriptsize}

\section{Details of supplementary files}
\label{app:suppl}
In the supplementary material we include two \textsc{mathematica} files 
(which can be obtained from the URL: \href{http://www.prl.res.in/~konar/data.html}
{http://www.prl.res.in/~konar/data.html} or from the source file in \texttt{arXiv})
where we spell out the details of the calculation of the unitarity constraints. 
In the file named \texttt{LRT\_Pot.nb} we construct the $2\to 2$ scattering matrices for
all the $q$-charged 2-particle states, mentioned in table~\ref{tab:2-particle-states}. 
One can also obtain the eigenvalues of those matrices by running that code by appropriately 
\textit{uncommenting} some \textit{commands}. We have collected all the independent eigenvalues 
of all the scattering matrices in the second file called \texttt{Eigenvalue\_collect.nb}. 

\bibliographystyle{JHEP}
\bibliography{0_LRT_unitarity_CB}

\end{document}